\documentclass[hidelinks,aps,pra,twocolumn,superscriptaddress,10pt]{revtex4-1} 

\usepackage{amsmath,amssymb}
\usepackage{mathrsfs}
\usepackage{epstopdf}
\usepackage{graphicx}
\usepackage{bm,color,bbm}
\usepackage{natbib}
\usepackage{hyperref}
\usepackage{caption}
\usepackage{subcaption}

\newcommand{\nn}{\nonumber}

\begin{document}


\title{Quantifying Tri-partite Entanglement with Entropic Correlations}


\author{James Schneeloch}
\email{james.schneeloch@gmail.com}
\affiliation{Air Force Research Laboratory, Information Directorate, Rome, New York, 13441, USA}

\author{Christopher C. Tison}
\affiliation{Air Force Research Laboratory, Information Directorate, Rome, New York, 13441, USA}

\author{Michael L. Fanto}
\affiliation{Air Force Research Laboratory, Information Directorate, Rome, New York, 13441, USA}
\affiliation{RIT Integrated Photonics group, Rochester Institute of Technology, Rochester, New York, 14623, USA}

\author{Shannon Ray}
\affiliation{Air Force Research Laboratory, Information Directorate, Rome, New York, 13441, USA}

\author{Paul M. Alsing}
\affiliation{Air Force Research Laboratory, Information Directorate, Rome, New York, 13441, USA}


\date{\today}

\begin{abstract}
We show how to quantify tri-partite entanglement using entropies derived from experimental correlations. We use a multi-partite generalization of the entanglement of formation that is greater than zero if and only if the state is genuinely multi-partite entangled. We develop an entropic witness for tripartite entanglement, and show that the degree of violation of this witness places a lower limit on the tripartite entanglement of formation. We test our results in the three-qubit regime using the GHZ-Werner state and the W-Werner state, and in the high-dimensional pure-state regime using the triple-Gaussian wavefunction describing the spatial  and energy-time entanglement in photon triplets generated in third-order spontaneous parametric down-conversion. In addition, we discuss the challenges in quantifying the entanglement for progressively larger numbers of parties, and give both entropic and target-state-based witnesses of multi-partite entanglement that circumvent this issue.
\end{abstract}


\maketitle

\section{Introduction}
When quantum objects (e.g., particles) have the capacity to interact, they have the capacity to become entangled. Any complex quantum system will have many interacting quantum elements, which makes a full understanding of these larger-scale systems challenging. Among other aspects, fully understanding these multi-partite systems requires fully understanding the forms of multi-partite (and multi-particle) entanglement \footnote{A state of $N$ parties is genuinely $N$-partite entangled if and only if the state cannot be obtained using entangled states of $N-1$ or fewer parties.} that exist between these parties, and has been intensely studied over the last two decades. With the advent of applied quantum information, device-independent secure quantum communication, and quantum computation, understanding and characterizing all forms of entanglement in real media is also an essential capability that must be mastered to make these technologies scalable toward quantum supremacy. Moreover, multipartite entanglement has been shown to give advantages in quantum networking protocols (e.g., quantum secret sharing \cite{PhysRevA.59.1829} and network key distribution \cite{Epping_2017}) as well as in quantum metrology \cite{toth2014quantum}. However, scalably characterizing multipartite entanglement for arbitrary states remains a open challenge. 

Entanglement between two parties has matured to the point that it is possible to experimentally quantify much of the entanglement present \cite{schneeloch2018quantifying,schneeloch2019record} without performing unscalable calculations (NP-hard in general) or an unfeasible determination of the complete quantum state (through tomography). Entanglement between three or more parties is still relatively underdeveloped. Though there exist multipartite entanglement monotones for pure states \cite{coffman2000distributed,ou2007monogamy,eisert2001schmidt, SzalayMultiEntMeasures}, adapting these measures for mixed states is challenging for both fundamental and practical reasons.

Fundamentally, the main barriers to quantifying (genuine) multipartite entanglement are two-fold. The first challenge comes from local operations and classical communication (LOCC) being the primary way to tell whether one state is more entangled than another \footnote{If quantum state $\hat{\rho}$ can be converted into state $\hat{\sigma}$ by LOCC operations, but not the other way around, then $\hat{\rho}$ is more entangled than $\hat{\sigma}$.}. While this works perfectly for ordering bipartite-entangled states, there are pairs of tripartite entangled states that cannot be converted into each other in either direction via LOCC \footnote{In, \cite{Contreras2019} it was shown that in spite of the incomparability via LOCC between different-multi-partite entangled states, one can identify the three-party gebit (i.e., the GHZ state) as the maximally multi-partite entangled state of three-qubits. This is accomplished by enlarging the definition of a multi-partite entanglement measure to be monotonic not just under LOCC operations, but under all operations that map the set of biseparable states onto itself.}. The second challenge is in developing a unit of multi-partite entanglement. While all bipartite entangled states can be synthesized using copies of two-qubit Bell states (i.e., ebits), the minimum set of entangled states needed to be able to synthesize all multi-partite states is unknown \cite{walter2016multipartite} (and remains an open problem). It is because of these issues that most multi-partite entanglement monotones are purely geometric, and that resource-based measures of multi-partite entanglement describing the number of elementary multi-partite states required or produced in quantum information protocols need further development.

In this paper, we develop a conservative bound to the tripartite entanglement of formation (denoted here as $E_{3\text{F}}$) discussed in \cite{SzalayMultiEntMeasures} using quantum entropies obtained from experimental correlations, as this measure has the greatest potential to be a resource-based measure with further theoretical development. We show how this strategy works both for three systems entangled in either discrete or continuous-variable degrees of freedom (e.g., photon triplets generated in third-order parametric down-conversion \cite{CoronaTripletSPDCPRA2011,Chang3OrdSPDCSupercond_PRX_2020}).

As a tripartite entanglement monotone: $E_{3\text{F}}$ is nonzero if and only if the state is genuinely tripartite entangled; it is invariant under local unitary transformations, and is non-increasing under LOCC. What distinguishes $E_{3\text{F}}$ from other tripartite entanglement measures is that it is additive under tensor products of pure states, so that $E_{3\text{F}}$ for $N$ copies of a pure state is $N$ times larger than for a single copy. This additivity is essential for a resource-based measure of entanglement.

The ordinary entanglement of formation, $E_{\text{F}}$, quantifies the number of ebits needed to synthesize a bipartite entangled state with LOCC, but it remains to be shown what protocol exemplifies $E_{3\text{F}}$. Even so, $E_{3F}$ has a principal unit of the three-qubit GHZ state (here called the three-party \emph{gebit} for three-party GHZ-entangled bit). Without an exemplary protocol, we can still say whether some high-dimensional tripartite state contains more tripartite entanglement than a number of gebits (see Appendix E for example), but understanding what the exemplary protocol is is an important subject for future research. Where multi-partite states can require multiple varieties of maximally entangled states to synthesize through LOCC, any single-valued resource measure of multi-partite entanglement is defined with respect to just one maximally multi-partite entangled state. Such measures may describe the number of copies necessary, but not sufficient to carry out a quantum information protocol.

\section{Foundations and Motivation: The tripartite Entanglement of Formation}
Entanglement is defined by departure from separability. If the joint quantum state of two parties $A$ and $B$ factors out into a product of states (one for $A$ and one for $B$), or is a mixture of such products, then that state is separable. Any state not separable is by definition entangled. The question of entanglement becomes more complex for three or more parties, because there are multiple ways a quantum state of three parties $A$, $B$, and $C$ can factor out into products of states contining fewer parties. As such, any meaningful discussion of entanglement in three or more parties requires pointing out from which classes of separable (or partially-separable) states the entanglement is being defined.

While the joint quantum state of $ABC$ factoring out into a triple product (one for $A$, one for $B$ and one for $C$) is fully separable (and so are mixtures of these products), a joint quantum state $ABC$ factoring out into only two products (say, one for $AB$ and one for $C$) is only bi-separable (as are mixtures of these states). Bi-separable states of $ABC$ can contain entanglement between two parties, but not between three. 

In the left-hand side of Fig.~1, we created a diagram of the different classes of tri-partite states, and the convex sets pertaining to different classes of separable and bi-separable states. To begin, each class of biseparable states forms a convex set (depicted as one of the three circles in the left diagram) because any mixture of two states in a given class produces a state in that same class. As an example, the class listed as $A\otimes BC$ represents the set of biseparable states that factor into a product of a state of $A$ and a joint state of $BC$. These sets intersect one another, and remarkably, there are states existing in the joint intersection of all three biseparability classes (the intersection given by the Rouleaux (curved) triangle) that are nonetheless not fully separable (the set depicted by the equilateral triangle of the same vertices).

While all non-fully-separable states are entangled in some fashion, non-bi-separable states of $ABC$ may or may not be genuinely tri-partite entangled. Measures of tripartite entanglement rule out not just all forms of separability in $ABC$, but also that the state of $ABC$ can be made out of any mixture of bi-separable states from different classes (here called biseparably-derived). Merely ruling out all forms of biseparability demonstrates full inseparability, but not genuine tri-partite entanglement. As an example of a fully inseparable state that is not tri-partite entangled, consider the three-way mixture $\hat{\rho}_{\text{insep.}}$:
\begin{align}\label{gogo}
\hat{\rho}_{\text{insep}}\!&=\!\frac{1}{3}\Big(\!|\Phi^{+}_{AB}\rangle\langle\Phi^{+}_{AB}|\otimes|0_{C}\rangle\langle0_{C}|\nn\\
&\qquad+|\Phi^{+}_{BC}\rangle\langle\Phi^{+}_{BC}|\otimes|0_{A}\rangle\langle0_{A}|\nn\\
&\qquad+|\Phi^{+}_{AC}\rangle\langle\Phi^{+}_{AC}|\otimes|0_{B}\rangle\langle0_{B}|\Big)\nn\\
&|\Phi^{+}_{AB}\rangle=\frac{1}{\sqrt{2}}\big(|0,0\rangle + |1,1\rangle\big)
\end{align}
This state is biseparably derived (by construction), while each party is entangled with the other two. Surprisingly, the reduced two-party state (tracing out any third party) is also separable, which makes this a state that is entangled, but where no two out of three parties are entangled, and which has no genuine tripartite entanglement.

In the right-hand side of the diagram, we have the corresponding separable sets of states when one traces out a third party. For example, the set $A\otimes B$ is the set of states $ABC$ that are separable when tracing over $C$ (or when only considering the reduced state $AB$). Here, it is clear that there exist genuinely tri-partite entangled states that are nonetheless separable when considering only two out of three parties (such as the GHZ state shown in \eqref{GHZstate}).  For a carefully, concisely, and clearly laid out discussion of the hierarchy of multi-parite entanglement and partial separability, see \cite{SzalayMultiEntMeasures}.

\begin{figure}[t]
\centerline{\includegraphics[width=\columnwidth]{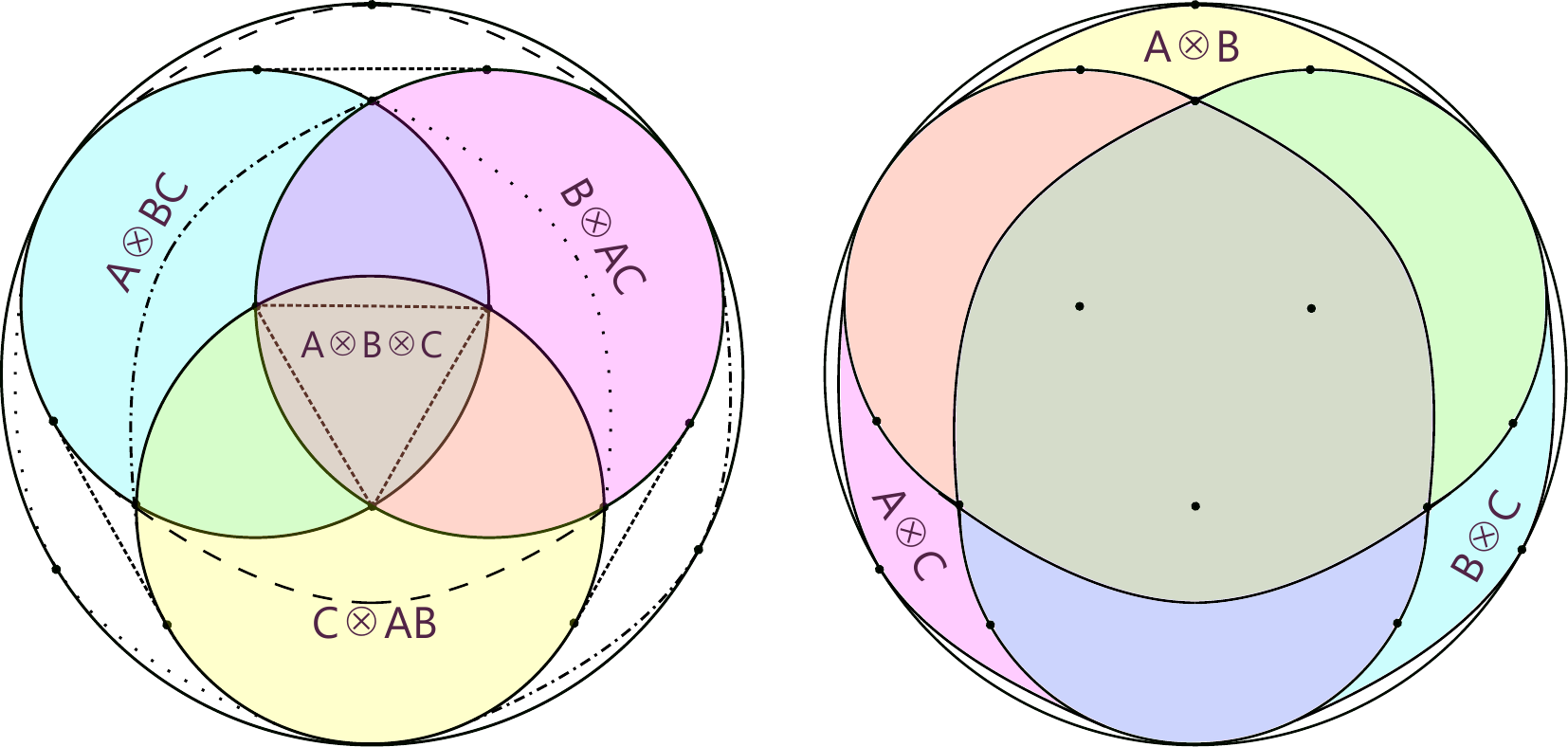}}
\caption{Left: Diagram of the tripartite separability classes of quantum states contained within the set of all tri-partite states. The three classes of bi-separable states contain the class of fully-separable states (straight-edged triangle), but there are states that are bi-separable all three ways (curved (Rouleaux) triangle) that are not fully separable. Conversely, there are mixtures of bi-separable states from different classes that are fully inseparable (not within any class), but which are still not genuinely tripartite-entangled. Right: Diagram of the bipartite separability classes within the set of tripartite quantum states. There are tripartite-entangled states whose bipartite subsystems are separable, and any bipartite separability class encompasses the convex hull of two tripartite separability classes.}
\end{figure}

\subsection{The Tripartite entanglement of formation}
The entanglement of formation $E_{\text{F}}(\hat{\rho})$ is one of the most ubiquitous measures of two-party entanglement \cite{WootersEntForm1998} because of its desirable properties. Like all entanglement measures, it satisfies the necessary axioms of entanglement monotonicity. It is invariant under local unitary transformations. It decreases monotonically under local operations and classical communication (LOCC), and it is zero for separable states. On top of this, $E_{\text{F}}$ is a faithful measure (being zero if and only if the state is separable), and it is additive under tensor products of pure states. As a resource measure, this means that $E_{\text{F}}$ for a product of different pure states is the sum of $E_{\text{F}}$ for each state individually. On top of this, $E_{\text{F}}$ is a popular measure because it has a simple physical interpretation in an exemplary protocol. Where an ebit is a maximally entangled 2-qubit state (i.e., a Bell state), $E_{\text{F}}(\hat{\rho})$ gives the number of ebits required on average to synthezise copies of $\hat{\rho}$ using LOCC, since LOCC alone cannot create entanglement.

In 2015, Szalay \cite{SzalayMultiEntMeasures} showed how one can generalize the entanglement of formation to be a measure defined with respect to any set of classes of separable states within the hierarchy of multipartite states. For a quantum state $\hat{\rho}_{ABC}$ of parties $A$, $B$, and $C$, the tripartite entanglement of formation $E_{3\text{F}}(ABC)$ is defined as:
\begin{equation}
E_{3\text{F}}(ABC)\equiv \min_{|\psi_{i}\rangle}\Big(\sum_{i}p_{i}\min\{S_{i}(A),S_{i}(B),S_{i}(C)\}\Big)
\end{equation}
where the first minimum is taken over all pure-state decompositions of $\hat{\rho}_{ABC}$. If the joint state of $ABC$ is already pure, $E_{3\text{F}}$ simplifies to the minimum between $S(A)$, $S(B)$, and $S(C)$, where for example $S(A)$ is the Von Neumann entropy of subsystem $A$. For general $N$-partite entanglement, the second minimum would be taken over the entropies of all subsystems formed in a bipartite split.

The tripartite entanglement of formation $E_{3\text{F}}$ is invariant under local unitary transformations, is non-increasing under LOCC, and is nonzero if and only if $\hat{\rho}_{ABC}$ is genuinely tripartite entangled (i.e., cannot be derived from mixing product states with kets containing two or fewer parties). What gives $E_{3\text{F}}$ potential as a resource-based measure is that like the ordinary entanglement of formation $E_{\text{F}}$, it is also additive under tensor products of pure states. This means that when using the three-party gebit as a principal unit of this aspect of three-party entanglement, $E_{3\text{F}}(\hat{\rho})$ gives the number of gebits that together would have the same amount of tripartite entanglement as $\hat{\rho}$. In contrast, the most ubiquitous measure of tripartite entanglement, the three-tangle \cite{coffman2000distributed} is not additive on products of pure states, and can be zero for some tripartite-entangled states, due to its being a measure of residual entanglement \footnote{Residual entanglement is the difference between the entanglement shared between a party $A$ and a group of parties $(B_{1},...,B_{N})$, and the entanglements $A$ shares with each subsystem of $B$ individually. As a concept for measuring tri-partite entanglement, it is only well-defined for pure states, as there are biseparably-derived mixed states (e.g., \eqref{gogo}) whose residual entanglement can be above zero.}.

\section{Bounding Tripartite Entanglement}
To obtain a conservative bound to $E_{3F}$, our approach has two steps. We first develop a relation between $E_{3F}$ and quantum conditional entropies, and then show how experimental correlations can bound these quantum entropies, bounding $E_{3F}$ in turn.

To begin, we point out that all quantum states of three parties $A$, $B$, and $C$ that are not genuinely tri-partite entangled must obey the relation:
\begin{equation}\label{QuantEntWitness}
S(A|BC) + S(B|AC) + S(C|AB) \geq -2\log(D_{\mathrm{max}}),
\end{equation}
where $D_{\text{max}}=\max\{D_{A},D_{B},D_{C}\}$ is the maximum dimension of parties $A$, $B$, and $C$, and $S(A|BC)=S(ABC)-S(BC)$ is the Von Neumann conditional entropy of subsystem $A$ on joint subsystem $BC$, of the tripartite state given by density matrix $\hat{\rho}_{ABC}$. See Appendix for proof. We prove that the amount of violation of this inequality both witnesses genuine tripartite entanglement, and provides a lower limit to $E_{3\text{F}}$.

To prove this relation bounds $E_{3\text{F}}$ from below, we begin with the amount of violation $V$ of \eqref{QuantEntWitness}:
\begin{equation}\label{E3min}
V\equiv -S(A|BC) - S(B|AC) - S(C|AB) -2\log(D_{\text{max}})\leq 0.
\end{equation}
Next, we use the fact that the quantum entropy (including conditional entropy) is concave, so that $V$ cannot decrease for any pure-state decomposition:
\begin{equation}
V\!\!\leq\!\sum_{|\psi\rangle_{i}}\!p_{i}\Big(\!-S_{i}(\!A|BC) - S_{i}(\!B|AC) - S_{i}(\!C|AB)\!\Big) -2\log(\!D_{\text{max}})
\end{equation}
and has no smaller value even for the minimal pure state decomposition:
\begin{equation}
V\leq \min_{|\psi\rangle_{i}}\Big\{\sum_{i}p_{i}\Big(S_{i}(A) + S_{i}(B) + S_{i}(C)\Big)\Big\} -2\log(D_{\text{max}}),
\end{equation}
where for a pure state $|\psi\rangle_{ABC}$, $-S(A|BC) = S(A)$.

Between $S_{i}(A)$, $S_{i}(B)$, and $S_{i}(C)$, the two largest of these three quantities are still each less than or equal to $\log(D_{\text{max}})$. The factors of $\log(D_{\text{max}})$ cancel out in $V$ to give us the result
 \begin{equation}
V\!\!\leq \!\min_{|\psi\rangle_{i}}\!\Big\{\!\!\sum_{i}\!p_{i}\!\Big(\!\!\min\{\!S_{i}(A),\!S_{i}(B),\!S_{i}(C)\!\}\!\!\Big)\!\!\Big\}=E_{3\text{F}}(ABC).
\end{equation}
As large as the difference between $(S_{i}(A),S_{i}(B),S_{i}(C))$ and $\log(D_{\text{max}})$ might be, this inequality is tight, in that there are bi-separable entangled states for which the two largest of $(S_{i}(A),S_{i}(B),S_{i}(C))$ actually equal $\log(D_{\text{max}})$, such as when two parties are in a maximally-entangled Bell state, while the third stands alone. 

Having shown how quantum conditional entropies provide a lower bound to $E_{3\text{F}}$, we next show how experimental correlations can bound these quantum entropies, which in turn, bound $E_{3\text{F}}$.

\subsection{Discrete observables}
Given a $D$-level quantum system, any pair of observables $\hat{Q}$ and $\hat{R}$ with respective sets of eigenstates $\{|q_{i}\rangle\}_{i=1}^{D}$ and $\{|r_{j}\rangle\}_{j=1}^{D}$ obey the entropic uncertainty relation \footnote{For a full review of entropic uncertainty relations and their applications, see \cite{ColesRMP}}:
\begin{align}\label{entUnvRel}
H(Q) + H(R) &\geq \log(\Omega) + S(\hat{\rho}),\\
:\;\Omega &\equiv \min_{i,j}\Bigg(\frac{1}{|\langle q_{i}|r_{j}\rangle|^{2}}\Bigg)\nn
\end{align}
where $H(Q)$ is the Shannon entropy of the measurement probabilities of the outcomes of $\hat{Q}$ given by $\text{Tr}\big[\hat{\rho}|q_{i}\rangle\langle q_{i}|\big]$:
\begin{equation}
H(Q)=-\sum_{i}\text{Tr}\big[\hat{\rho}|q_{i}\rangle\langle q_{i}|\big]\log\Big(\text{Tr}\big[\hat{\rho}|q_{i}\rangle\langle q_{i}|\big]\Big),
\end{equation}
and $S(\hat{\rho})$ is the von Neumann entropy of the quantum state $\hat{\rho}$. The value $\Omega$ ranges between $1$ and $D$, where the minimum value is obtained when $\hat{Q}$ and $\hat{R}$ commute (and there is a simultaneously determined pair of eigenstates where the inner product $\langle q_{i}|r_{j}\rangle=1$). The maximum value is attained when $\hat{Q}$ and $\hat{R}$ are maximally uncertain with respect to one another (and the inner product $\langle q_{i}|r_{j}\rangle\rightarrow1/\sqrt{D}$ for all $i$ and $j$). Here and throughout this paper, all logarithms are taken to be base-$2$, since we measure information in bits.

The entropic uncertainty principle can be used to place an upper limit on the quantum entropy of that system. To bound quantum \emph{conditional} entropy, we can use the uncertainty principle in the presence of quantum memory \cite{Berta2010} adapted for classical entropies and for three parties:
\begin{equation}\label{DiscEntCorr}
H(Q_{A}|Q_{B},\!Q_{C}) + H(R_{A}|R_{B},\!R_{C})\!\geq\! \log(\Omega) \!+\! S(\!A|BC).
\end{equation}
Expressed differently, this inequality is given as:
\begin{equation}
-S(\!A|BC)\!\geq\! \log(\Omega)\!-\!\Big(\!\!H(Q_{A}|Q_{B},\!Q_{C}) \!+\! H(R_{A}|R_{B},\!R_{C})\!\!\Big).
\end{equation}
Thus, using the experimental correlations between $(Q_{A},Q_{B},Q_{C})$ and $(R_{A},R_{B},R_{C})$, one can find minimum values for $-S(A|BC)$, $-S(B|AC)$ and $-S(C|AB)$, which when added together with $-2\log(D_{\text{max}})$, gives a minimum value for $E_{3F}$. 

For pure states, one can bound $E_{3\text{F}}$ more tightly using just the minimum value between $-S(A|BC)$, $-S(B|AC)$ and $-S(C|AB)$. As long as the correlations are strong enough that the lower bounds to each of these quantites are greater than zero, the minimum of these bounds is a lower bound to $E_{3\text{F}}$.

\subsection{Continuous observables}
For continuous observables, we can bound the quantum conditional entropies like $S(A|BC)$ in a similar fashion as for discrete observables. Using the methods in \cite{schneeloch2018quantifying}, one can derive a corresponding entropy constraint for continuous observables that are Fourier conjugates of one another (e.g., position/momentum, energy/time, field quadratures, etc). For position $x$ and momentum $k=p/\hbar$, we have the relation 
\begin{equation}
h(x_{A}|x_{B},x_{C}) + h(k_{A}|k_{B},k_{C})\geq \log(2\pi) + S(\!A|BC), 
\end{equation}
where for example, $h(x_{A}|x_{B},x_{C})$ is the continuous Shannon entropy of probability density $\rho(x_{A},x_{B},x_{C})$ conditioned on variables $x_{B}$ and $x_{C}$. To use a relation like this experimentally, one can use the discrete probabilities that come from coarse-graining $x$ and $k$ into bins of size $\Delta x$ and $\Delta k$, respectively. Because coarse-graining cannot decrease entropy, we can use these discrete probabilities in the relation:
\begin{equation}
H\!(\!X_{\!A}|X_{\!B},\!X_{\!C}\!) + H(\!K_{\!A}|K_{\!B},\!K_{\!C})\!\geq\! \log\!\!\Big(\!\frac{2\pi}{\Delta x\Delta k}\!\Big) + S(\!A|BC) 
\end{equation}
to establish a minimum value to $-S(A|BC)$. However, because the relation \eqref{QuantEntWitness} between these quantum entropies and $E_{3\text{F}}$ explicitly includes the dimension $D$, we can only bound $E_{3\text{F}}$ from below for continuous-variable systems if the underlying state is known to be pure. It may be that tighter bounds exist if $S(ABC)$ is known, since a small deviation from a pure state should result in a small change in the threshold for tripartite entanglement, but more research is needed. This relation is still useful in theoretical tests where the various entanglements of formation of an infinite-dimensional pure tripartite state cannot yet be computed (which is almost every case). In Section III.D, we show how we can estimate $E_{3\text{F}}$ for the triple-Gaussian wavefunction of photon-triplets produced in third-order spontaneous parametric down-conversion.

\subsection{Test1: The GHZ-Werner state and W state}
In order to test the strength of our bound on $E_{3\text{F}}$, we evaluated it for a GHZ-Werner state of $3$-qubits defined as:
\begin{equation}
\hat{\rho}_{\text{GW}}\equiv p\;|\operatorname{GHZ}_{3}\rangle\langle \operatorname{GHZ}_{3}| + (1-p) \hat{\rho}_{\text{MM}}
\end{equation}
where
\begin{equation}\label{GHZstate}
|\operatorname{GHZ}_{3}\rangle\equiv \frac{1}{\sqrt{2}}\Big(|0\rangle^{\otimes 3} + |1\rangle^{\otimes 3}\Big),
\end{equation}
$p$ is the GHZ mixing fraction between zero and one, and $\hat{\rho}_{\text{MM}}$ is the maximally mixed state of three qubits. Because the GHZ-Werner state has such a simple form, we were able to plot as a function of $p$ both the exact value of $V$ (the lower bound to $E_{3\text{F}}$ from \eqref{E3min}) and the lower bound to $V$ using \eqref{DiscEntCorr} with observables $\hat{Q}$ and $\hat{R}$ given as the Pauli $\hat{\sigma}_{x}$ and $\hat{\sigma}_{z}$.

\begin{figure}[t]
\centerline{\includegraphics[width=0.85\columnwidth]{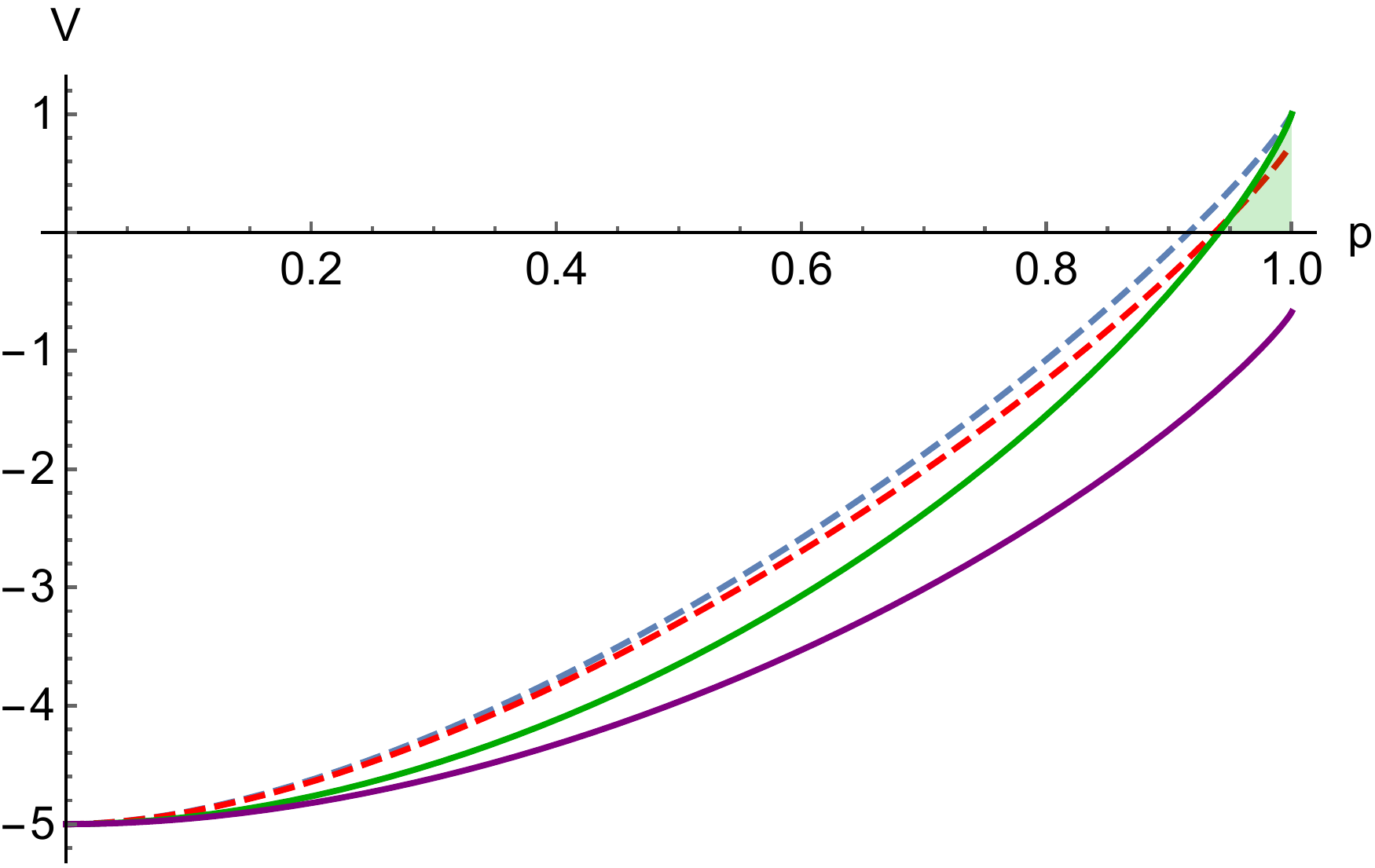}}
\caption{Plot of $V$, the minimum value of $E_{3\text{F}}$ of the pure state fraction $p$. The red and blue dashed curves give the exact values of $V$ for the W-Werner state and the GHZ-Werner state, respectively, while the purple and green solid curves give the respective lower bounds to $V$ using the probabilities obtained from $\sigma_{x}$ and $\sigma_{z}$ correlations. The green shaded region where $p>0.9406$ is where three-party entanglement is quantified with the GHZ-Werner state.}
\end{figure}

Here, we see for $p$ greater than $0.94$, we reach the threshold for quantifying tripartite entanglement, where the largest lower bound to $E_{3\text{F}}$ is unity at $p=1$, which is also the exact value of $E_{3\text{F}}$ at that value of $p$. Though there are smaller values of $p$ sufficient to witness tripartite entanglement (e.g., using the fidelity to the GHZ state shows tripartite entanglement for $p>\frac{3}{7}$ \cite{guhne2009entanglement}), these fidelity-based witnesses did not (at the time) give any information about how much multipartite entanglement exists in the system (though we show one such strategy in Section IV.A.). Actually quantifying tripartite entanglement gives us more information about the resources present.

In addition to the GHZ-Werner state, we also examined how our lower bound to $E_{3\text{F}}$ behaves for the three-qubit W-Werner state:
\begin{equation}
\hat{\rho}_{\text{WW}}\equiv p\;|W_{3}\rangle\langle W_{3}| + (1-p) \hat{\rho}_{\text{MM}}
\end{equation}
where
\begin{equation}
|W_{3}\rangle\equiv \frac{1}{\sqrt{3}}\Big(|0,0,1\rangle +|0,1,0\rangle + |1,0,0\rangle\Big),
\end{equation}
$p$ is the $W$ mixing fraction between zero and one, and $\hat{\rho}_{\text{MM}}$ is the maximally mixed state of three qubits. For all values of $p$, we found no violation. However, if we instead evaluate the bound directly by calculating the quantum conditional entropies in \eqref{E3min} (that may be determined through quantum state tomography), we can still quantify as much $E_{3\text{F}}$ as $0.7549$ gebits (where explicit calculation of $E_{3\text{F}}$ for $|W\rangle$ gives $\approx0.9183$ gebits).

\subsection{Test 2: High-dimensional tri-partite entanglement of photon triplets}
\subsubsection{Spatial entanglement}
For photon triplets generated in third-order degenerate collinear spontaneous parametric down-conversion (TODC-SPDC), the momentum wavefunction in one spatial dimension is given by:
\begin{align}
\psi&(k_{1},k_{2},k_{3})=N \psi_{\text{pump}}(k_{1}+k_{2}+k_{3})\times\nn\\
&\times \operatorname{sinc}\Big(\frac{3 L_{z}}{4 k_{p}}\big((k_{1}+k_{2})^{2} + (k_{2}+k_{3})^{2} + (k_{3} + k_{1})^{2}\big)\Big).
\end{align} 
where here, we simplify $(\vec{k}_{1})_{x}$ as $k_{1}$ to simplify notation. For a full derivation, see Appendix B. The sum of squares of sums of momentum inside the $\operatorname{sinc}$ function is a quadratic form, and can be expressed in terms of rotated orthogonal momentum coordinates:
\begin{align}
(k_{1}&+k_{2})^{2} + (k_{2}+k_{3})^{2} + (k_{3} + k_{1})^{2}\nn\\
&= k_{u}^{2} + k_{v}^{2} + 4 k_{w}^{2},
\end{align} 
where $k_{u}=\frac{1}{\sqrt{6}}(2 k_{1}-k_{2}-k_{3})$. $k_{v}=\frac{1}{\sqrt{2}}(k_{2}-k_{3})$, and $k_{w}=\frac{1}{\sqrt{3}}(k_{1}+k_{2}+k_{3})$. Since this $\operatorname{sinc}$ function approximately factors into a product of $\operatorname{sinc}$ functions for each orthogonal component, and because we can approximate the $\operatorname{sinc}$ function with a Gaussian function \cite{Schneeloch_SPDC_Reference_2016} (as well as the pump), we eventually arrive at the following form for the triple-Gaussian wavefunction for third-order SPDC:
\begin{align}
\psi(k_{u},k_{v},k_{w})=N &\operatorname{exp}[-\frac{8a}{9}k_{u}^{2}]\times \operatorname{exp}[-\frac{8a}{9}k_{v}^{2}]\nn\\
&\times \operatorname{exp}[-(3 \sigma_{p}^{2} + \frac{32 a}{9})k_{w}^{2}]
\end{align} 
where $a\equiv\frac{3 L_{z}\lambda_{p}}{8\pi n_{p}}$. Here: $L_{z}$ is the length of the nonlinear medium; $\lambda_{p}$ is the wavelength of the pump light; $n_{p}$ is the index of refraction of the nonlinear medium at the pump wavelength; $\sigma_{p}$ is one quarter of the $1/e^{2}$ beam diameter of the pump beam, and $N$ is a normalization constant.

The spatial multipartite entanglement in third-order SPDC is a consequence of strong momentum correlations due to momentum conservation with the pump, along with near-perfect position correlations due to the triphoton being created at one point in space. Because the correlations are between three parties, the uncertainty principle actually forbids near-perfect correlations \footnote{Near-perfect correlations in this context is where knowledge of one variable allows the determination of all others to a high degree of precision. The probability density of three nearly-perfectly-correlated variables approximately traces out a curve in that three-dimensional space.} in both position and momentum (See Appendix D for details) even though this is not the case for two parties.

\begin{figure*}[t]
\centerline{\includegraphics[width=0.9\textwidth]{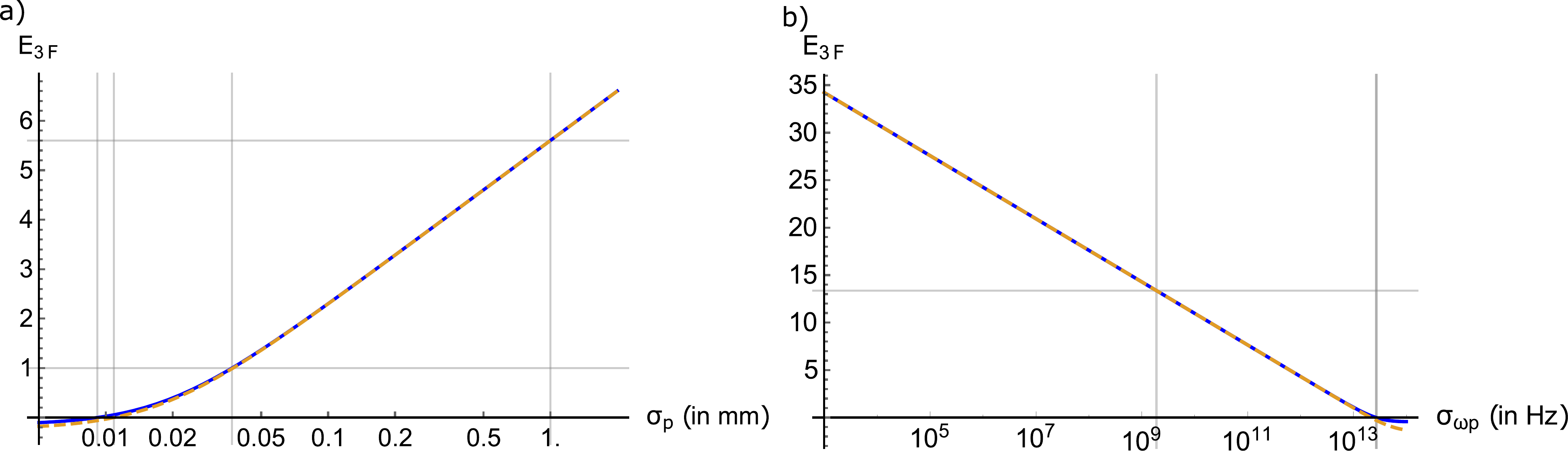}}
\caption{a) Plot of the minimum value of the tripartite entanglement of formation $E_{3\text{F}}$ (obtained from conditional position and momentum entropies of the Triple Gaussian approximation) of an entangled photon triplet generated in degenerate, collinear third-order SPDC. While the solid curve gives the exact lower bound to $E_{3\text{F}}$, the dashed curve is the approximation $E_{3\text{F}}\geq \log(\frac{e}{2}\frac{\sigma(x_{A}-x_{B})}{\sigma(x_{A})})+0.2075$, where the constant is independent of $a$ and $\sigma_{p}$ (see Appendix D1 for derivation); $\sigma(x_{A})=\sqrt{\sigma_{p}^{2}+\frac{16a}{9}}$, and $\sigma(x_{A}-x_{B})=\sqrt{\frac{16 a}{9}}$. Vertical guidelines indicate the pump widths of the zero intercepts of $E_{3\text{F}}$ and its approximation, followed by the pump widths $0.0370$ mm corresponding to $E_{3\text{F}}=1$ gebit, and $1.0$ mm corresponding to $E_{3\text{F}}=5.6000$ gebits, respectively. b) Plot of $E_{3\text{F}}$ in the frequency-time degree of freedom for the photon triplet source as a function of the pump bandwidth parameter $\sigma_{\omega_{p}}$. The dashed trendline (see Appendix D for derivation) is the bound $E_{3\text{F}}\geq -\log(e\sigma(t_{A}-t_{B})\sigma(\omega_{A}+\omega_{B}+\omega_{C}))+0.207519$, where the constant is independent of $\sigma_{\omega_{p}}$ and $b$; $\sigma(t_{A}-t_{B})=\sqrt{\frac{16b}{9}}$; $\sigma(\omega_{A}+\omega_{B}+\omega_{C})=1/\sqrt{4(\frac{8b}{27}+\frac{1}{4\sigma_{\omega_{p}}^{2}})}$.Vertical guidelines indicate $\sigma_{\omega_{p}}=1.94$GHz, and $27.3$THz for pump bandwidth of a He-Cd gas laser, and the marginal bandwidth (nearly equal to the zero intercept), respectively. The horizontal guideline gives the value $E_{3\text{F}}$ of $13.37$ gebits.}
\end{figure*}

Using the triple-Gaussian wavefunction, and its Fourier transform for both position and momentum statistics, we are able to evaluate all of the entropies necessary to place a lower limit to $E_{3\text{F}}$. As an example, we consider a 10 mm-long crystal of Aluminum Nitride \cite{LuChi32018}, a nonlinear material with a bandgap of $\approx6$eV (transparent from about $200$nm to $>\!15\mu$m), and a third-order nonlinear susceptibility ($\chi^{(3)}$) of approximately $160\text{pm}^{2}/V^{2}$ (or nonlinear index of about $2.3\times10^{-19}\text{m}^{2}/W$). If we assume a $325$nm pump (with index $n_{p}=2.247$) and beam radius $\sigma_{p}=1.0$mm, then we find $E_{3\text{F}}$ is no less than approximately $5.6000$ gebits, or more tripartite entanglement than can be supported on a 16-qubit state space (See Fig.~3a for plot). 

\subsubsection{Energy-time entanglement}
In the limit of a narrowband pump, there is even more tripartite entanglement to be had in third-order SPDC. This is especially fortunate, as single-mode nonlinear optical waveguides can greatly enhance the efficiency of photon triplet generation by maintaining a high intensity over the entire length of the nonlinear medium. The joint frequency amplitude is given by the integral:
\begin{equation}
\psi(\omega_{1}\!,\omega_{2},\omega_{3})\!=\!N\!\!\int \!d\omega_{p}\psi_{\text{pump}}(\omega_{p})\delta(\Delta\omega) \operatorname{sinc}\Big(\frac{L_{z}\Delta k_{z}}{2}\Big).
\end{equation} 
Expanding, $\Delta k_{z}$ to second order in frequency, and simplifying in the same fashion as was used to obtain the spatial wavefunction, we obtain:
\begin{align}
\psi(\omega_{1},\omega_{2},\omega_{3})&=N\psi_{\text{pump}}(\omega_{w}\sqrt{3}) \times\nn\\
&\times \operatorname{sinc}\Big(\frac{L_{z}\kappa}{4}\big(\omega_{u}^{2}+\omega_{v}^{2} + \omega_{w}^{2}\big)\Big).
\end{align} 
where $\omega_{u}=\frac{1}{\sqrt{6}}(2 \omega_{1}-\omega_{2}-\omega_{3})$. $\omega_{v}=\frac{1}{\sqrt{2}}(\omega_{2}-\omega_{3})$, $\omega_{w}=\frac{1}{\sqrt{3}}(\omega_{1}+\omega_{2}+\omega_{3}-\omega_{p0})$, and $\omega_{p0}$ is a constant equal to the center frequency of the pump light. In addition, $b\equiv L_{z}\kappa/4$ and $\kappa$ is the group velocity dispersion $(d^{2}k/d\omega^{2})$ at one third of the pump frequency. As with the case in spatial entanglement, we use these orthogonal frequency coordinates and approximate the joint frequency amplitude as a product of three Gaussian functions, one in each coordinate:
\begin{align}
\psi(\omega_{u},\omega_{v},\omega_{w})=N &\operatorname{exp}[-\frac{8b}{9}\omega_{u}^{2}]\times \operatorname{exp}[-\frac{8b}{9}\omega_{v}^{2}]\nn\\
&\times \operatorname{exp}[-(\frac{3}{4\sigma_{\omega_{p}}^{2}} + \frac{8 b}{9})\omega_{w}^{2}]
\end{align}
Using this joint frequency amplitude to describe the light in the same setup where we previously quantified spatial entanglement, we find (see Fig.~3b) that there is just as much potential for very large amounts of three-party entanglement as has been shown for two-party entanglement in the energy-time degree of freedom of ordinary second-order SPDC \cite{schneeloch2018quantifying}. Using the same experimental example as we did for spatial entanglement, and assuming the relatively broad bandwidth of a 325nm He-Cd gas laser $\sigma(\omega_{p})\approx 1.9$GHz,(and $\kappa=1.01\times10^{25}s^{2}/m$) we predict approximately $13.37$ gebits, or more tripartite entanglement than a $40$-qubit state can support. The amount is only limited by how narrow one can make the bandwidth of the pump light, which is readily available at MHz- or even kHz-scale bandwidths with current technology.

As with many other photon triplet sources, it is a subject of ongoing research \cite{corona2011experimental,huang2013generation} to increase the triplet generation rate to make them more feasible. Knowing how much tripartite entanglement can be present in these degrees of freedom, we see that there is much to be excited about in the resources contained in these states.

\section{Discussion: Challenges in quantifying multipartite entanglement}
Properly quantifying tripartite entanglement for mixed states is a formidable challenge for a number of reasons. For two parties, entanglement can be quantified in terms of ebits (or Bell pairs), which, with LOCC can be used to synthesize any two-party state. For three parties however, there is no known Minimum Reversible Entanglement Generating Set (MREGS) of entangled states from which all three-party states can be synthesized via LOCC. Even including Bell pairs for every bipartite subsystem and for every bipartite split of the tripartite system (i.e., the set of maximally entangled states with repsect to every separability class) is still not sufficient to synthesize every tripartite state \cite{acin2003structure}. While this set of states can generate any tripartite state within any bi-separable class via LOCC, there are fully inseparable states that cannot be synthesized in this way. In addition, the practical aspects of scalably quantifying multi-partite entanglement present a greater challenge than for bipartite entanglement. The difficulty in computing multi-partite entanglement measures is at least NP-hard (the difficulty in determining entanglement in general), and efficient methods to quantify multi-partite entanglement of mixed states in high-dimensional (Continuous-variable) degrees of freedom remain to be developed.

That said, we have shown how experimental correlations can be used to bound the amount of tripartite entanglement (as measured by $E_{3\text{F}}$) from below. We have shown its effectiveness both for low-dimensional systems, and for high-dimensional continuous-variable systems. An intriguing open question is whether similar entropic bounds exist for $N$-partite generalizations of the entanglement of formation ($E _{N\text{F}}$ for $N>3$). Tripartite entanglement is special because there are only three bipartite splits of three parties. For $N$ partites, there are $(2^{N-1}-1)$ bipartite splits, making the number of measurements grow exponentially with $N$. However, one can use entropic correlations to witness multi-partite entanglement in a scalable way (though the fidelity to the $N$-partite GHZ state remains optimal \cite{guhne2009entanglement}) if the correlations are strong enough. For $N$ parties correlated in observables $\hat{Q}$ and $\hat{R}$, violating the inequality
\begin{equation}\label{StrongMultiRel}
\sum_{i=1}^{N}\!\!\Big(\!\!H(Q_{i}|Q_{i+1}) \!+\! H(R_{i}|R_{i+1},...,R_{i+(N-1)})\!\Big)\!\geq\! 2\log(\Omega),
\end{equation}
(where the sum wraps around $N$ to $1$ again) will witness genuine $N$-partite entanglement, and is maximally violated with an $N$-partite GHZ state (See Appendix C for details). Better still, if the state to be measured is sufficiently close to a target state (e.g., the  $N$-partite GHZ state), one can measure a small number of density matrix elements to efficiently quantify as well as qualify its multi-partite entanglement.

\subsection{Quantifying multi-partite entanglement close to target states with individual density matrix elements}
Although our paper shows how experimental correlations can quantify $E_{3F}$ with minimal prior knowledge of the state, we have recently found in the literature that there exist highly-efficient solutions to quantify entanglement when the state being measured is close to a known state. While a sufficiently high fidelity to a particular multi-partite-entangled target state witnesses multi-partite entanglement \cite{guhne2009entanglement}, one can use similar custom-tailored bounds to also quantify the multi-partite entanglement as well. 

In \cite{WuMultiEnt}, Wu \emph{et~al} showed how specific elements of a density matrix can be used to efficiently bound from below a measure of genuine $N$-partite entanglement based on the linear entropy $S_{\text{L}}(\hat{\rho})=2[1-\text{Tr}[\hat{\rho}^{2}]]$. Their measure given here as $E_{N\text{L}}$, is given by:
\begin{equation}
E_{N\text{L}}(\hat{\rho})=\min_{|\psi\rangle}\sum_{i}p_{i}\min_{\gamma}\sqrt{S_{L}(\hat{\rho}_{\gamma i})},
\end{equation}
where $\gamma$ in an index running over all possible subsystems formed by tracing over one side of a bipartite split, and $\hat{\rho}_{\gamma i}$ is the density operator of the $\gamma$ subsystem of the $i$-th pure state in the decomposition of the full $N$-party state. They also point out how $E_{N\text{L}}(\hat{\rho})$ is related to $E_{N\text{F}}(\hat{\rho})$ through convexity arguments and relations between different entropy functions.

The function $-\log(1-x^{2}/2)$ is a convex function of $x$. The quantum collision entropy $S_{C}(\hat{\rho})$ is $-\log_{2}(\text{Tr}[\hat{\rho}^{2}])$, and is a direct function of the linear entropy:
\begin{equation}
S_{C}(\hat{\rho})=-\log_{2}\Big(1-\frac{\sqrt{S_{L}(\hat{\rho})}^{2}}{2}\Big).
\end{equation} 
Because of this convexity, and that $E_{N\text{L}}(\hat{\rho})$ is a mean value of $\sqrt{S_{\text{L}}}$, we can say the following. Given a bound $\mathcal{B}\leq E_{N\text{L}}$, it follows that:
\begin{equation}
-\log_{2}\Big(1-\frac{\mathcal{B}^{2}}{2}\Big)\leq E_{N\!C}(\hat{\rho})\leq E_{N\text{F}}(\hat{\rho}),
\end{equation}
where
\begin{equation}
E_{N\!C}(\hat{\rho})=\min_{|\psi\rangle}\sum_{i}p_{i}\min_{\gamma}S_{C}(\hat{\rho}_{\gamma i}),
\end{equation}
and the last inequality bounding $E_{N\text{F}}$ comes from $S_{C}(\hat{\rho})$ being less than or equal to the von Neumann entropy of the same density matrix. To construct this bound $\mathcal{B}$, we refer the reader to \cite{WuMultiEnt} and \cite{MaEntBound}, but here we give the specific example well-adapted to the $N$-qubit GHZ state:
 \begin{align}
 \mathcal{B}&=2|\langle 0^{\otimes N}|\hat{\rho}|1^{\otimes N}\rangle|\nn\\
 &-\!\!\sum_{q=1}^{2^{N}-2}\!\!\!\sqrt{\langle q|\hat{\rho}|q\rangle\langle 2^{N}\!-\!1-q|\hat{\rho}|2^{N}\!-\!1-q\rangle}.\label{BoundB}
 \end{align}
To explain the compact notation with an example, $q=5$ expressed in binary is $101$, and the four-qubit ket $|q\rangle$ for $q=5$ is given by $|0,1,0,1\rangle$. This bound $\mathcal{B}$ is as tight as the fidelity-based witnesses in \cite{guhne2009entanglement}, having an equal tolerance for noise in the GHZ-Werner state for $N$ qubits (e.g., $p>3/7$ implies $\mathcal{B}>0$ and $E_{3\text{F}}>0$ for the three-qubit  $\hat{\rho}_{\text{GW}}$.), while providing the additional information of a minimum nonzero value for the entanglement measure. However, this bound works well only for states close to the $N$-qubit GHZ state. If one instead inputs the W state, the bound $\mathcal{B}<0$ and no multi-partite entanglement could be quantified (though bounds constructed especially for states close to the W state work well). 

Interestingly, the bound $\mathcal{B}$ adapted to states close to the $N$-qubit GHZ state can be further improved without losing its robustness to noise. The second term in \eqref{BoundB} is expressible as an inner product between two vectors. Using the Cauchy-Schwarz inequality, and the fact that summing over all probabilities gives unity leaves us with the result:
 \begin{align}
 \mathcal{B}&\geq2|\langle 0^{\otimes N}|\hat{\rho}|1^{\otimes N}\rangle|\nn\\
 &\,\,\,+|\langle 0^{\otimes N}|\hat{\rho}|0^{\otimes N}\rangle| +|\langle 1^{\otimes N}|\hat{\rho}|1^{\otimes N}\rangle| -1
 \end{align}
In other words, by summing the magnitudes of just the four corner-elements of the $N$-qubit density matrix, any sum greater than unity witnesses genuine $N$-partite entanglement, and the amount by which it is greater than unity is a lower bound for $\mathcal{B}$, which in turn gives a lower bound for $E_{N\text{F}}$. This restricted bound is equally robust to noise, and is equally optimized for states close to the GHZ state (and even maintains the critical value $p\geq3/7$ for the $\text{GHZ}_{3}$ - Werner state). For high-dimensional quantum systems, where maximally correlated systems still have a comparatively large number of significant elements of the density matrix, research into efficiently quantifying multipartite entanglement is ongoing.

\section{Conclusion}
In our investigations, we have developed a relation that allows us to bound from below the tripartite entanglement of formation $E_{3F}$, both for general 3-qudit systems (where each system can have independent dimension), and for pure-state continuous-variable systems. In doing so, we found that photon triplets generated in third-order spontaneous parametric down-conversion have several three-party gebits (i.e., 3-qubit-GHZ states) worth of tripartite entanglement both in their spatial and especially in their energy-time degrees of freedom. In doing this, we have come across fundamental challenges in quantifying multi-partite entanglement, where a single number is insufficient to characterize all forms of entanglement present in a multi-partite state. However, this may yet be resolved. While gebits and all bipartite ebits are not sufficient to synthesize every tripartite state, they also cannot be converted into each other by LOCC, and represent resources in their own right. Futher developing $E_{3F}$ and other resource measures of multi-partite entanglement may shed much light on this issue.

\begin{acknowledgments}
We gratefully acknowledge support from the Air Force Office of Scientific Research LRIR 18RICOR028 and LRIR 18RICOR079, as well as insightful discussions with Ms. Laura Wessing, Dr. A. Matthew Smith, and Dr. Dimitri Uskov. Any opinions, findings and conclusions or recommendations expressed in this material are those of the author(s) and do not necessarily reflect the views of AFRL.
\end{acknowledgments}

\bibliography{EPRbib17}

\newpage

\appendix

\begin{widetext}
\section{Proof of tripartite entanglement witness (Equation (3)}
To prove the relation that all biseparably derived states obey:
\begin{equation}\label{a1}
S(A|BC) + S(B|AC) + S(C|AB) \geq -2\log(D_{\text{max}}),
\end{equation}
we consider the case that the state is biseparable with respect to the partition $A\otimes BC$.
For this class of states, $S(A|BC)$ simplifies to $S(A)$, which must be greater than or equal to zero. Then, $S(B|AC)$ and $S(C|AB)$ simplify to $S(B|C)$ and $S(C|B)$, respectively, which are each bounded from below by $-\log(D_{\text{max}})$ (such as with maximum entanglement between $B$ and $C$), where $D_{\text{max}}$ is the maximum dimension between systems $A$, $B$, or $C$. This proves that all states biseparable with respect to the bipartition $A\otimes BC$ obey the relation.

Next, because the relation \eqref{a1} is symmetric under all permutations of $A$, $B$, or $C$, all states biseparable with respect to any bipartion ($A\otimes BC$, $B\otimes AC$ or $C\otimes AB$) obey this relation, so that violation witnesses full inseparability.

Finally, we use the fact that the quantum conditional entropy is a concave function. Because of this, any state that is a mixture of biseparable states from any class (known as biseparably derived) must also obey the relation. Therefore, violation of this relation witnesses genuine tripartite entanglement.

\section{Derivation of the spatial wavefunction for third-order degenerate collinear SPDC}
Just as ordinary second-order SPDC is described by the Hamiltonian \cite{schneeloch2019introduction}:
\begin{equation}
\hat{H}_{NL}=\frac{1}{3}\int d^{3}r\big(\zeta_{ij\ell}^{(2)}(\vec{r}) \hat{D}_{i}^{+}(\vec{r},t)\hat{D}_{j}^{-}(\vec{r},t)\hat{D}_{\ell}^{-}(\vec{r},t) + \text{H.C.}\big),
\end{equation}
third-order Spontaneous parametric down-conversion is described by the nonlinear-optical Hamiltonian:
\begin{equation}
\hat{H}_{NL}=\frac{1}{4}\int d^{3}r\big(\zeta_{ij\ell m}^{(3)}(\vec{r}) \hat{D}_{i}^{+}(\vec{r},t)\hat{D}_{j}^{-}(\vec{r},t)\hat{D}_{\ell}^{-}(\vec{r},t)\hat{D}_{m}^{-}(\vec{r},t) + \text{H.C.}\big).
\end{equation}
Here, $\zeta_{ij\ell m}^{(3)}(\vec{r})$ is the third-order inverse nonlinear-optical susceptibility tensor, and:
\begin{equation}
\hat{D}^{-}(\vec{r},t)=-i\sum_{\vec{k},s}\sqrt{\frac{\epsilon_{0}n_{\vec{k}}^{2}\hbar\omega_{\vec{k}}}{2V}}\hat{a}_{\vec{k},s}^{\dagger}\vec{\epsilon}_{\vec{k},s}e^{i (\vec{k}\cdot \vec{r}-\omega_{\vec{k}}t)}
\end{equation}
is the operator for the negative frequency component of the electric displacement field. In addition, $\hat{a}_{\vec{k},s}^{\dagger}$ is the creation operator for a photon of momentum indexed by $\vec{k}$ and polarization indexed by $s$, with polarization direction $\vec{\epsilon}_{\vec{k},s}$, and quantization volume $V$.

Next, we are able to use first-order time-dependent perturbation theory to describe the state of the down-converted light because of its relatively low probability. Given that a large scale experiment might have as much of $10$ meters of optical path length, a pair generation rate of $30$ million per second would still have no more than one pair in the experiment at any given time. Photon triplet sources have not yet reached this generation rate, so it is a valid approximation to use the first order.

Since we are only interested in the spatially varying aspect of third-order SPDC, we lump together all constants that do not vary spatially into the normalization constant $N$:
\begin{equation}
|\psi_{3SPDC}\rangle \approx N\sum_{\vec{k}_{p},\vec{k}_{1},\vec{k}_{2},\vec{k}_{3}}\int dt\int d^{3}r \chi_{eff}^{(3)}(\vec{r})e^{i\Delta\vec{k}\cdot\vec{r}}e^{-i\Delta\omega t}\hat{a}_{\vec{k}_{p}}\hat{a}_{\vec{k}_{1}}^{\dagger}\hat{a}_{\vec{k}_{2}}^{\dagger}\hat{a}_{\vec{k}_{3}}^{\dagger}|\alpha_{\vec{k}_{p}},0,0,0\rangle
\end{equation}
Here, the state is given as an operator acting on the initial state of the field,which is the vacuum state at the down-converted frequencies and a coherent state at the pump frequencies.

Next, we use the paraxial approximation to separate the transverse and longitudinal dependence of the pump, signal, and idler fields. The integral over $dt$ enforces $\Delta\omega\approx 0$, and the integrals over $x$ and $y$ where the nonlinear medium is larger than the pump beam enforces transverse momentum conservation: $\Delta k_{x}\approx 0\approx \Delta k_{y}$. Where the pump beam's coherence is much longer than the nonlinear medium of length $L_{z}$ and susceptibility $\chi_{\text{eff}}^{(n)}$, the integral over $z$ is a finite $\operatorname{sinc}$ function, and would be over-simplified by approximating to a delta function. With these steps, the state of the down-converted field (pump omitted) simplifies to:
\begin{equation}
|\psi_{3SPDC}\rangle \approx N\sum_{\vec{q}_{1},\vec{q}_{2},\vec{q}_{3}}\alpha_{p}(\vec{q}_{1}+\vec{q}_{2}+\vec{q}_{3}) \operatorname{sinc}\Big(\frac{\Delta k_{z}L_{z}}{2}\Big)\hat{a}_{\vec{q}_{1}}^{\dagger}\hat{a}_{\vec{q}_{2}}^{\dagger}\hat{a}_{\vec{q}_{3}}^{\dagger}|0,0,0\rangle
\end{equation}
where $\vec{q}=\vec{k}-k_{z}\hat{z}$ is the projection of $\vec{k}$ onto the $xy$ plane, and $\alpha_{p}(\vec{q})$ is the amplitude of the pump field at transverse momentum $\vec{q}$. To continue the simplification, we take the continuum limit, replacing sums by integrals, and converting discrete momentum creation operators into continuous momentum creation operators:
\begin{equation}
|\psi_{3SPDC}\rangle \approx N\int d^{2}q_{1}d^{2}q_{2}d^{2}q_{3}\Psi(\vec{q}_{1},\vec{q}_{2},\vec{q}_{3})\hat{a}_{\vec{q}_{1}}^{\dagger}\hat{a}_{\vec{q}_{2}}^{\dagger}\hat{a}_{\vec{q}_{3}}^{\dagger}|0,0,0\rangle
\end{equation}
with the triphoton wavefunction given by:
\begin{equation}
\Psi(\vec{q}_{1},\vec{q}_{2},\vec{q}_{3})\equiv \alpha_{p}(\vec{q}_{1}+\vec{q}_{2}+\vec{q}_{3}) \operatorname{sinc}\Big(\frac{\Delta k_{z}L_{z}}{2}\Big)
\end{equation}
With this, we need to express $\Delta k_{z}$ in terms of the transverse momenta $\vec{q}$.
\begin{figure}[h]
\centerline{\includegraphics[width=0.65\textwidth]{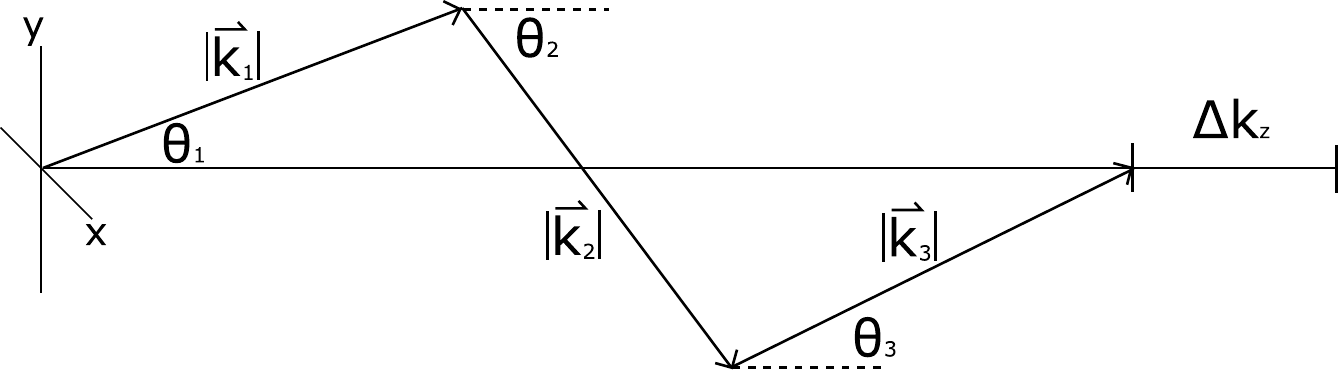}}
\caption{Diagram of the momentum vectors adding up to the pump momentum vector aligned in the $z$ direction.}
\end{figure}

To understand how the longitudinal momentum offset $\Delta k_{z}$ corresponds to the transverse momenta of the daughter photons, we point to Figure 4. In collinear third-order SPDC, the transverse components of the daughters' momenta must add to zero, while the longitudinal components sum to a total that differs from the pump by amount $\Delta k_{z}$. Since we assume approximately degenerate third-order SPDC, the magnitude of the momenta of the daughter photons are equal to each other, and at one third of the pump momentum. These specifications, give us the longitudinal relation:
\begin{equation}
\Delta k_{z}=\frac{k_{p}}{3}\Big(3-\cos(\theta_{1})-\cos(\theta_{2})-\cos(\theta_{3}) \Big)
\end{equation}
which in the small angle approximation becomes:
\begin{equation}
\Delta k_{z}\approx\frac{k_{p}}{3}\Big(\frac{\theta_{1}^{2}}{2} + \frac{\theta_{2}^{2}}{2} + \frac{\theta_{3}^{2}}{2} \Big)
\end{equation}
Then, the transverse relation simplifies in a similar way:
\begin{equation}
\theta_{1}\approx\frac{|\vec{q}_{1}|}{k_{1}}=\frac{|\vec{q}_{2} + \vec{q}_{3}|}{k_{1}}=\frac{3}{k_{p}}|\vec{q}_{2} + \vec{q}_{3}|
\end{equation}
giving us the triphoton wavefunction:
\begin{equation}
\Psi(\vec{q}_{1},\vec{q}_{2},\vec{q}_{3})\equiv \alpha_{p}(\vec{q}_{1}+\vec{q}_{2}+\vec{q}_{3}) \operatorname{sinc}\Big(\frac{3L_{z}\lambda_{p}}{8\pi n_{p}}\big(|\vec{q}_{1} + \vec{q}_{2}|^{2}+|\vec{q}_{2} + \vec{q}_{3}|^{2}+|\vec{q}_{3} + \vec{q}_{1}|^{2}\big)\Big)
\end{equation}
The final approximation, giving the triphoton spatial wavefunction in this paper is that the $\operatorname{sinc}$ function is approximately separable into a product of two $\operatorname{sinc}$ functions, one for each dimension (i.e., $\operatorname{sinc}(x^{2}+y^{2}) \approx \operatorname{sinc}(x^{2}) \operatorname{sinc}(y^{2})$), which is valid both for small values of the $\operatorname{sinc}$ function, and for its later approximation as a Gaussian function.

\section{Proof of entropic criterion for N partite entanglement}
Our entropic criterion for $N$-partite entanglement is given by the cyclic sum of entropies:
\begin{equation}
\sum_{i=1}^{N}\!\!\Big(\!\!H(Q_{i}|Q_{i+1}) \!+\! H(R_{i}|R_{i+1},...,R_{i+(N-1)})\!\Big)\!\geq\! 2\log(\Omega),
\end{equation}
where the index $i+(N-1)$ wraps around from $N$ to $1$, when $i$ is sufficiently large.

Any biseparably-derived state is a mixture of pure biseparable states of different classes. From this, we find the following:
\begin{equation}\label{MultiRelImpLong}
\sum_{i=1}^{N}\!\!\Big(\!\!H(Q_{i}|Q_{i+1}) \!+\! H(R_{i}|R_{i+1},...,R_{i+(N-1)})\!\Big)\geq \sum_{j} p_{j}\;\sum_{k} q_{jk}\Bigg(\sum_{i=1}^{N}\!\!\Big(\!\!H(Q_{i}|Q_{i+1}) \!+\! H(R_{i}|R_{i+1},...,R_{i+(N-1)})\!\Big)\Bigg)_{jk}
\end{equation}
where the sum over $j$ is for each class of biseparable states, and for $k$, the pure states within each class. Because we are summing over cyclic permutations, any bipartite split must eliminate two correlations instead of one (see Fig.~5).

\begin{figure}[h]
\centerline{\includegraphics[width=0.3\textwidth]{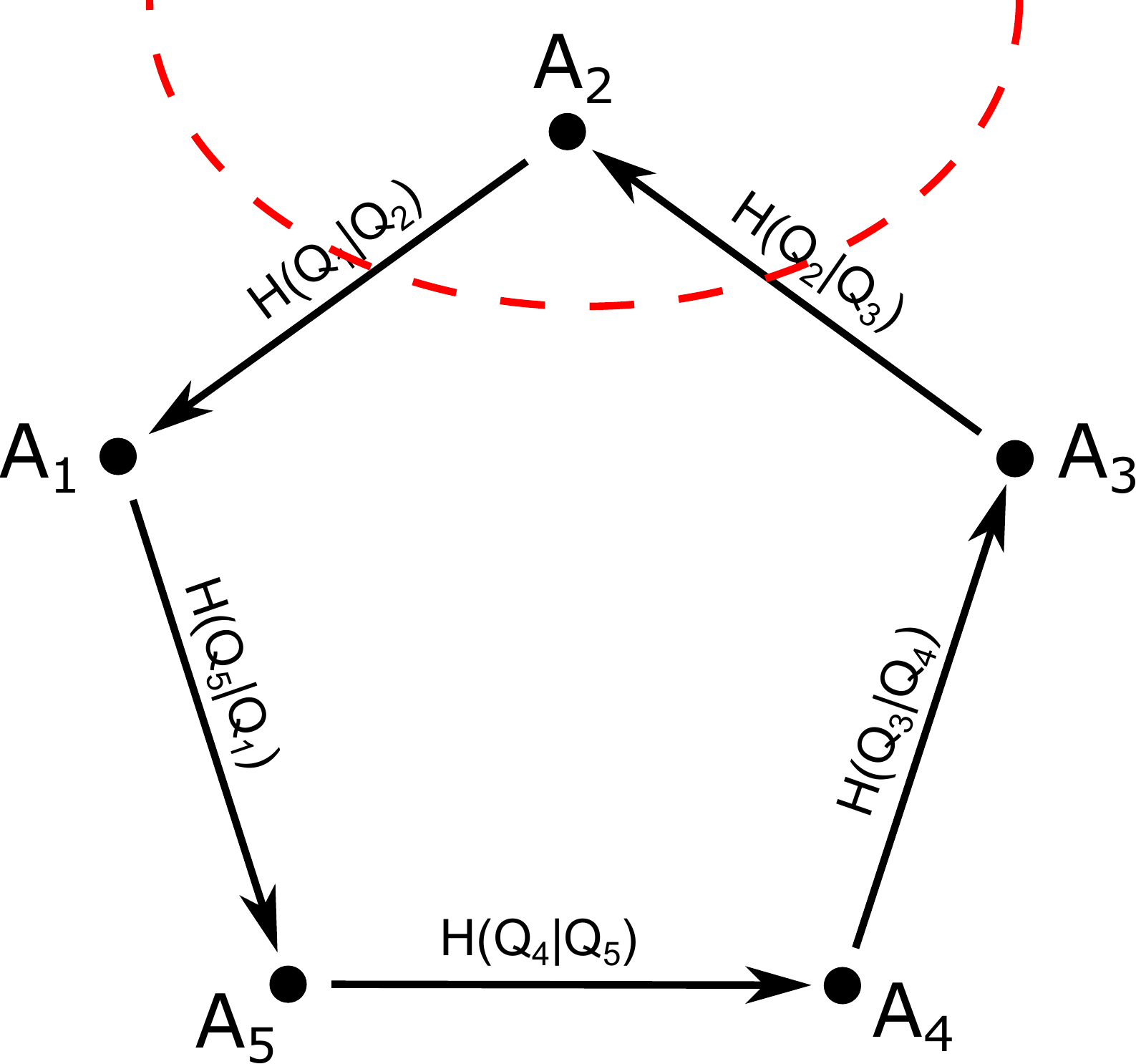}}
\caption{Diagram of graph expressing correlation structure of our $N$-partite entanglement criterion (here, $N=5$ as an example). By using a cycle graph, we see that any cut has to eliminate at least two edges (correlations).}
\end{figure}

Because of this, any split must for two values of $i$ reduce $H(Q_{i}|Q_{i+1})$ to $H(Q_{i})$. The rest of the proof follows:
\begin{align}
\sum_{i=1}^{N}\!\!\Big(\!\!H(Q_{i}|Q_{i+1}) \!+\! H(R_{i}|R_{i+1},...,R_{i+(N-1)})\!\Big)&\geq \sum_{j} p_{j}\;\sum_{k} q_{jk}\Bigg(\sum_{i=1}^{N}\!\!\Big(\!\!H(Q_{i}|Q_{i+1}) \!+\! H(R_{i}|R_{i+1},...,R_{i+(N-1)})\!\Big)\Bigg)_{jk}\nn\\
&\geq \sum_{j} p_{j}\;\sum_{k} q_{jk}\Bigg(\sum_{i}^{2 \text{ values}}\!\!\Big(\!\!H(Q_{i}) \!+\! H(R_{i}|R_{i+1},...,R_{i+(N-1)})\!\Big)\Bigg)_{jk}\nn\\
&\!\!\!\!\!\!\!\!\!\!\!\!\!\!\!\!\!\!\!\!\!\!\!\!\!\!\!\!\!\!\!\!\!\!\!\!\!\!\!\!\geq \sum_{j} p_{j}\;\sum_{k} q_{jk}\Bigg(\sum_{i}^{2 \text{ values}}\!\!\Big(\!\!H(Q_{i}|R_{i+1},...,R_{i+(N-1)}) \!+\! H(R_{i}|R_{i+1},...,R_{i+(N-1)})\!\Big)\Bigg)_{jk}\nn\\
&\geq \sum_{j} p_{j}\;\sum_{k} q_{jk}\Bigg(\sum_{i}^{2 \text{ values}}\!\!\Big(\log(\Omega)\Big)_{jk}\Bigg)\nn\\
&\geq\sum_{j} p_{j}\;\sum_{k} q_{jk}\Big(2\log(\Omega)\Big)\nn\\
&=2\log(\Omega)
\end{align}

\subsection{Side note: Extension to complete graphs}
Any cut of a cyclic graph with $N$ vertices eliminates at least two edges. The entropic criterion we developed with this uses $N$ pairs of entropies and has a bound of $2\log(\Omega)$. Any cut of a complete graph with $N$ vertices eliminates at least $N-1$ edges. If we were to develop a multipartite entropic criterion that uses all pairs of correlations, we would need $N(N-1)/2$ pairs of entropies, but it would give a bound of $(N-1)\log(\Omega)$. For states that are symmetric under exchange of parties, the sum of entropies and the bound would both increase by the same factor of $N-1$, giving no advantage to summing over all pairwise correlations.

\section{Approximation to continuous-variable entanglement formulas}
In Figs.~3a and 3b, we were able to plot $E_{3\text{F}}$ as a function of the pump width, and the pump bandwidth respectively because for a Gaussian wavefunction, the solution was analytic. The trendlines given as logarithms of ratios of different widths were first discovered and later derived, as seen in Appendix Section D.1.

\begin{figure}[h]
\centerline{\includegraphics[width=0.55\textwidth]{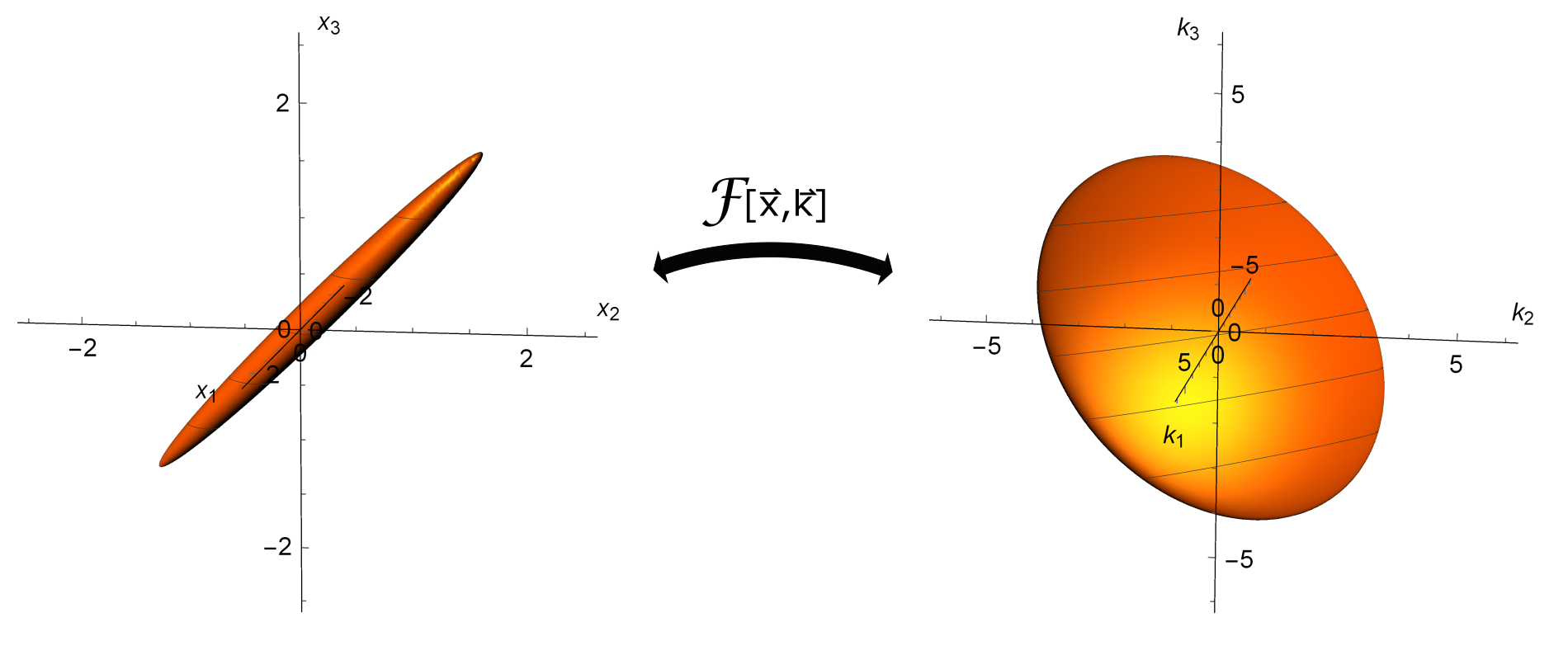}}
\caption{Diagram of triple-Gaussian wavefunction $\psi(x_{1},x_{2},x_{3})$. In position, $x_{1}\approx x_{2}\approx x_{3}$, and the resulting wavefunction is narrow along the two principal coordinates $x_{u}$ and $x_{v}$, but longer in $x_{w}$. In momentum space, the biphoton wavefunction is Fourier-transformed, becoming wide in $k_{u}$ and $k_{v}$, but narrow in $k_{w}$.}
\end{figure}

The reasoning behind these trendlines is that for ordinary SPDC, the double-Gaussian wavefunction $\psi(x_{1},x_{2})$ has an entanglement bounded by the logarithm of its Schmidt number $K$, \cite{Schneeloch_SPDC_Reference_2016} which is given as the ratio of the marginal width $\sigma(x_{1})$ over the correlated width $\sigma(x_{1}|x_{2})\approx\sigma(x_{1}-x_{2})$. Using the logarithm of the corresponding ratio for the triphoton wavefunction yields excellent if not exact agreement. This makes sense because the relationship between correlated quantities before and after a Fourier transform is different between three parties than between two. 

For two parties, a long skinny double-Gaussian wavefunction has two principal widths with a given ratio between them. After transforming between position and momentum,  long widths in position transform to short widths in momentum, and the wavefunction remains long and skinny, albeit in a different direction. For three parties, there are three principal widths that get transformed in the same way. Because of this, a triphoton wavefunction cannot remain long and skinny in both position and momentum. Instead, prolate correlations in position transform to oblate correlations in momentum, and vise versa (see Fig.~ 6). As described in detail in Appendix Section D.2, the uncertainty principle limits the strength of correlations between three or more pairs of conjugate observables, even though it does not for two.

\subsection{Derivation of formulas for entanglement lower bounds of third-order SPDC pure-state wavefunction}
In position and momentum, the entanglement of the pure triphoton wavefunction (because it is also symmetric between parties) is bounded by the formula:
\begin{equation}
E_{3\text{F}}\geq-S(A|BC)\geq\log(2\pi)-(h(x_{A}|x_{B},x_{C})+h(k_{A}|k_{B},k_{C}))
\end{equation}
These quantum conditional entropies are generally difficult to calculate explicitly, but can be bounded effectively with simpler forms:
\begin{align}
h(x_{A}|x_{B},x_{C})&\leq h(x_{A}|x_{B})=h(x_{A}-x_{B}|x_{B})\leq h(x_{A}-x_{B})\leq \frac{1}{2}\log(2\pi e \sigma(x_{A}-x_{B})^{2})\\
h(k_{A}|k_{B},k_{C})&\leq h(k_{A}) \leq \frac{1}{2}\log(2\pi e \sigma(k_{A})^{2})\leq\frac{1}{2}\log\Big(\frac{\pi e}{2\sigma(x_{A})^{2}}\Big)
\end{align}
The successively looser bounds are accomplished from three properties. First, removing conditioning variables cannot decrease entropy. Second, for a constant variance $\sigma^{2}$, there is a maximum value to the continuous entropy given by a Gaussian distribution of the same variance. Third, the momentum variance $\sigma(k_{A})^{2}$ is bounded above by $1/(4\sigma(x_{A})^{2})$ through the Heisenberg uncertainty principle. Combining these results, we have that for the pure state wavefunction:
\begin{equation}
E_{3\text{F}}\geq-\log\Bigg(\frac{e}{2}\frac{\sigma(x_{A}-x_{B})}{\sigma(x_{A})}\Bigg)
\end{equation}
Using these same techniques, one can derive a similar bound for the entanglement present in the energy-time degree of freedom, where:
\begin{align}
h(t_{A}|t_{B},t_{C})&\leq h(t_{A}|t_{B})=h(t_{A}-t_{B}|t_{B})\leq h(t_{A}-t_{B}) \leq \frac{1}{2}\log(2\pi e \sigma(t_{A}-t_{B})^{2})\\
h(\omega_{A}|\omega_{B},\omega_{C})&=h(\omega_{A}+\omega_{B}+\omega_{C}|\omega_{B},\omega_{C})\leq h(\omega_{A}+\omega_{B}+\omega_{C})\leq \frac{1}{2}\log(2\pi e \sigma(\omega_{A}+\omega_{B}+\omega_{C})^{2})
\end{align}
gives us:
\begin{equation}
E_{3\text{F}}\geq-\log(e\sigma(t_{A}-t_{B})\sigma(\omega_{A}+\omega_{B}+\omega_{C}))
\end{equation}
When the pure state triphoton wavefunction is not symmetric between parties, these bounds may still be employed, but the minimum must be taken over all permutations of parties to truly bound the tripartite entanglement present.

\subsection{Side note: Optimum multipartite correlations are not perfect}
The notion that the uncertainty principle prevents conjugate observables from being perfectly correlated when considering three or more parties, was first discussed in \cite{Epping_2017}, where Epping showed algebraically that $N$ qubits perfectly correlated in one set of observables cannot be perfectly correlated, even pairwise, in the conjugate set. Using the entropic uncertainty principle in the presence of quantum memory, adapted for classical measurements of multiple parties, one can make this relationship more general and transparent for arbitrary conjugate observables $\hat{Q}$ and $\hat{R}$:
\begin{equation}
H(\!Q_{1},\!...,\!Q_{N-1}\!|Q_{N}\!) + H(\!R_{1},\!...,\!R_{N-1}\!|R_{N}\!) \geq \!(N-1)\!\log(D) + S(A_{1},...,A_{N-1}|A_{N})\geq \!(N-2)\!\log(D).
\end{equation}
Here, we see that in the limit that $\hat{Q}$ is perfectly correlated (so that $Q_{N}$ determines all of $(Q_{1},...,Q_{N-1})$ and $H(\!Q_{1},\!...,\!Q_{N-1}\!|Q_{N}\!)=0$, $\hat{R}$ is nowhere near so well correlated, and $ H(\!R_{1},\!...,\!R_{N-1}\!|R_{N}\!)\geq(N-2)\!\log(D)$. The best possible correlation in $\hat{R}$ can be shown by rearranging the relation:
\begin{equation}
H(Q_{1},Q_{2},...,Q_{N-1}|Q_{N}) + H(R_{1}|R_{2},...,R_{N}) \geq \!(N-2)\!\log(D) - H(R_{2},...,R_{N-1}|R_{N})\geq 0.
\end{equation}
In other words, if one variable $Q_{i}$ determines the rest of the set $(Q_{1},...,Q_{N})$, then at best, all but one variable in the set $(R_{1},...,R_{N})$ is needed to determine the rest of the set $(R_{1},...,R_{N})$.

\section{Equivalance between two 3-party gebits and one 4-level 3-party GHZ state}
Given two, $3$-party gebits shared by $3$ parties $(A_{1},B_{1},C_{1},A_{2},B_{2},C_{2})$, where the sub-index indicates the first or the second copy of the subsystem sent to $A$, $B$, or $C$, the joint state $|\psi\rangle$ is given by:
\begin{align}
|\psi\rangle&=\Bigg(\frac{1}{\sqrt{2}}\big(|0,0,0\rangle + e^{i\phi_{1}} |1,1,1\rangle\big)\Bigg)\otimes \Bigg(\frac{1}{\sqrt{2}}\big(e^{i\phi_{2}}|0,0,0\rangle + e^{i\phi_{3}}|1,1,1\rangle\big)\Bigg)\nn\\
&=\frac{1}{2}\Big(|0,0,0\rangle\otimes|0,0,0\rangle+e^{i\phi_{1}}|0,0,0\rangle\otimes|1,1,1\rangle+e^{i\phi_{2}}|1,1,1\rangle\otimes|0,0,0\rangle+e^{i\phi_{3}}|1,1,1\rangle\otimes|1,1,1\rangle\Big)\nn\\
&=\frac{1}{2}\Big(|0,0,0,0,0,0\rangle+e^{i\phi_{1}}|0,0,0,1,1,1\rangle+e^{i\phi_{2}}|1,1,1,0,0,0\rangle+e^{i\phi_{3}}|1,1,1,1,1,1\rangle\Big).
\end{align}
Something interesting occurs when we re-group the indices by party from $(A_{1},B_{1},C_{1},A_{2},B_{2},C_{2})$ to $(A_{1},A_{2},B_{1},B_{2},C_{1},C_{2})$. We obtain the state: 
\begin{align}
|\psi\rangle^{'}&=\frac{1}{2}\Big(|0,0,0,0,0,0\rangle+e^{i\phi_{1}}|0,1,0,1,0,1\rangle+e^{i\phi_{2}}|1,0,1,0,1,0\rangle+e^{i\phi_{3}}|1,1,1,1,1,1\rangle\Big)\nn\\
&=\frac{1}{2}\Big(|0,0\rangle\otimes|0,0\rangle\otimes|0,0\rangle+e^{i\phi_{1}}|0,1\rangle\otimes|0,1\rangle\otimes|0,1\rangle+e^{i\phi_{2}}|1,0\rangle\otimes|1,0\rangle\otimes|1,0\rangle+e^{i\phi_{3}}|1,1\rangle\otimes|1,1\rangle\otimes|1,1\rangle\Big),
\end{align}
which is a GHZ-state of three $4$-level systems (albeit with each $4$-level system being made out of a product state of two parties). Where the arbitrary phases $(\phi_{1},\phi_{2},\phi_{3})$ can only be identified when all parties cooperate, the utility of GHZ states in quantum secret sharing protocols is self-evident. It is similarly straightforward to show that the three-party entanglement of $n$ $3$-party gebits is equivalent to one, $2^{n}$-level 3-party GHZ state, both formally, and in their capacity of quantum secret sharing.

\end{widetext}

\end{document}